# Heteroepitaxy of Large-Area, Monocrystalline Lead Halide Perovskite Films on Gallium Arsenide

Deying Kong[1], Yu Zhang[2], Dali Cheng[3], Enze Wang[1], Kaiyuan Zhang[3], Huachun Wang[3], Kai Liu[1], Lan Yin[1,*] and Xing Sheng[3,*]

**Affiliations**

[1]School of Materials Science and Engineering, The Key Laboratory of Advanced Materials of Ministry of Education, State Key Laboratory of New Ceramics and Fine Processing, Center for Flexible Electronics Technology, Tsinghua University, Beijing, China

[2]Department of Physics, Tsinghua University, Beijing, China

[3]Department of Electronic Engineering, Beijing National Research Center for Information Science and Technology, Center for Flexible Electronics Technology, IDG/McGovern Institute for Brain Research, Tsinghua University, Beijing, China

*Correspondence to: lanyin@tsinghua.edu.cn; xingsheng@tsinghua.edu.cn

**ABSTRACT**

Lead halide perovskite materials have been emerging as promising candidates for high-performance optoelectronic devices. Significant efforts have sought to realize monocrystalline perovskite films at a large scale. Here, we epitaxially grow monocrystalline methylammonium lead tribromide (MAPbBr$_3$) films on lattice-matched gallium arsenide (GaAs) substrates at a centimeter scale. In particular, a solution-processed lead(II) sulfide (PbS) layer provides a lattice-matched and chemical protective interface for the solid-gas reaction to form MAPbBr$_3$ films on GaAs. Structure characterizations identify the crystal orientations in the trilayer MAPbBr$_3$/PbS/GaAs epi-structure and confirm the monocrystalline nature of MAPbBr$_3$ on PbS/GaAs. The dynamic evolution of surface morphologies during the growth indicates a two-step epitaxial process. These fundamental understandings and practical growth techniques offer a viable guideline to approach high-quality perovskite films for previously inaccessible applications.

**MAIN TEXT**

In the past few years, hybrid lead halide perovskites have become promising candidates as new generation semiconductors for advanced optoelectronic materials and devices, including remarkable developments like high-efficiency solar cells and light-emitting diodes (LEDs)[1, 2, 3, 4, 5]. In spite of these achievements, these devices are mostly based on polycrystalline perovskites, and it is imperative to obtain large-area, single crystalline perovskite films with a high quality on par with traditional semiconductors like silicon (Si) or gallium arsenide (GaAs)[6, 7, 8, 9]. Certain accomplishments have been achieved, for example, by growing monocrystalline perovskite films via lattice-matched homoepitaxy or metamorphic heteroepitaxy on uncommon substrates like monocrystalline perovskites[10, 11, 12, 13], metal halides[14, 15] or lead chalcogenides[16, 17]. Other attempts include the growth of perovskite crystals on more common but lattice-mismatched substrates like Si, GaN, quartz, gold, sapphire, etc.[18, 19, 20, 21, 22, 23], via spin coating, inkjet printing, mechanical slicing, space-confined crystallization, chemical vapor deposition, electrodeposition, etc[20, 22, 23, 24, 25, 26]. However, crystal sizes formed on these lattice-mismatched substrates are limited to scales ranging from several micrometers to several millimeters. In a word, existing approaches either require lattice-matched crystals that are not readily available, or utilize lattice-mismatched substrates that constrain the growth of large-area, monocrystalline perovskite films. In other words, the limited availability of large-area, lattice-matched substrates and their associated growth strategies hinder the formation of large-area, monocrystalline perovskite films, which has long been desired for unprecedented device applications.

In this work, we report the epitaxial growth of centimeter-scale, monocrystalline methylammonium lead tribromide ($CH_3NH_3PbBr_3$, or $MAPbBr_3$) perovskite films on lattice-matched GaAs wafers. Coincidentally, cubic-phase $MAPbBr_3$ has a lattice constant ($a$ = 5.91 Å)[27] close to that of zinc-blende GaAs ($a$ = 5.65 Å), making the lattice-matched or metamorphic epitaxy possible. To evaluate this hypothesis, we first attempt to grow MAPbBr$_3$ via conventional chemical vapor deposition (CVD)[28], following the reaction of methylammonium bromide (CH$_3$NH$_3$Br, or MABr) and lead(II) bromide (PbBr$_2$) at 350 °C (Figure S1a):

$$CH_3NH_3Br(g) + PbBr_2(g) = CH_3NH_3PbBr_3(s)$$

This reaction produces polycrystalline MAPbBr$_3$ on lattice-mismatched Si (100) substrates ($a$ = 5.43 Å) (Figure S1b), which is expected and in accordance with previous reports[28]. However, the same synthesis process results in the formation of lead (Pb) powders on GaAs (100), which makes the direct CVD growth of MAPbBr$_3$ unsuccessful (Figure S1c). This observation is likely to be attributed to more favorable reactions between bromides and GaAs at 350 °C [29, 30]:

$$3PbBr_2(g) + GaAs(s) = 3Pb(s) + GaBr_3(g) + AsBr_3(g)$$

$$6CH_3NH_3Br(g) + GaAs(s) = 6CH_3NH_2(g) + 3H_2(g) + GaBr_3(g) + AsBr_3(g)$$

which produce volatile gallium(III) bromide (GaBr$_3$) and arsenic tribromide (AsBr$_3$) with boiling points of 278 °C and 221 °C, respectively. In fact, many group III and group V halides form volatile gases (for example, boiling points are 201 °C for GaCl$_3$, 130 °C for AsCl$_3$, 173 °C for PBr$_3$, etc.). Therefore, even if the lattice matching requirement is satisfied, such detrimental processes could prevent the direct epitaxy of various halide perovskites on GaAs at elevated temperatures.

To bypass this challenge, we introduce a lead(II) sulfide (PbS) interface layer to ameliorate the epitaxial growth of MAPbBr$_3$ on GaAs (Figure 1). The PbS layer is grown on GaAs via chemical bath deposition (CBD) by the reaction of lead and sulfur ions in the alkaline solution (Figure 1a)[31, 32]:

$$Pb^{2+}(aq) + S^{2-}(aq) = PbS(s)$$

Subsequently, the MAPbBr$_3$ film is formed by the solid-gas reaction between MABr and PbS in the CVD tube at 145 °C (Figure 1b):

$$3CH_3NH_3Br(g) + PbS(s) = CH_3NH_3PbBr_3(s) + 2CH_3NH_2(g) + H_2S(g)$$

Such a reaction was previously employed for perovskite nanocrystals or quantum dots[33, 34, 35], but has not been explored to form large-area monocrystalline perovskite films. The introduction of the PbS interlayer is advantageous in multiple aspects: (1) PbS has a face-centered cubic (FCC), rock salt structure almost perfectly lattice-matched ($a$ = 5.93 Å) to MAPbBr$_3$, working as an ideal buffer layer between MAPbBr$_3$ ($a$ = 5.91 Å) and GaAs ($a$ = 5.65 Å); (2) PbS protects GaAs against detrimental reactions with halides; (3) PbS serves as a Pb source and directly reacts with MABr, reducing the CVD temperature from 350 °C to 145 °C. Such material choices and process strategies eventually cause the formation of large-area, monocrystalline MAPbBr$_3$ on GaAs, which is systematically analyzed subsequently.

Figure 2 presents materials and crystal characterizations for the PbS/GaAs and the MAPbBr$_3$/PbS/GaAs epitaxial structures. Figures 2a and 2b show that both the PbS and the MAPbBr$_3$ films form uniform coatings on a GaAs (100) wafer sample with a dimension of around 2 × 2 cm$^2$. In addition, the MAPbBr$_3$ film presents a highly-ordered island structure with cube-shaped crystals all oriented along the cleavage directions ([110] and [1$\bar{1}$0]) of GaAs (100) wafers (Figure 2b, bottom). Cross-sectional scanning electron microscopic (SEM) images (Figure 2c) and corresponding element mappings (Figure S2) indicate that the as-prepared PbS film has a thickness of ~500 nm after ~5 min CBD, and the MAPbBr$_3$ layer is ~200 nm thick after ~8 hour CVD. Electron Back Scatter Diffraction (EBSD) mappings (Figure 2d) clearly illustrate that both the PbS and the MAPbBr$_3$ films have monocrystalline characteristics, with a preferential growth in [110] and [100] directions, respectively. In agreement with EBSD data, out-of-plane x-ray diffraction (XRD) patterns in Figure 2e confirm that the PbS and the MAPbBr$_3$ films have growth planes of (220) and (100), respectively. For comparison, the same procedures to grow PbS and MAPbBr$_3$ films are performed on Si and glass substrates. As expected, XRD results indicate that these films are polycrystalline on Si and glass (Figure S3).

Pole figures in Figure 2f further elucidate in-plane monocrystalline structures for PbS and MAPbBr$_3$ films. The (111) pole figure of PbS with two spots separated azimuthally by 180° at a tilt angle of 35°, corresponding to the angle between (111) and (220) planes, indicates the in-plane orientation of PbS with respect to GaAs (Figure 2f, top). Additionally, the (211) pole figure of MAPbBr$_3$ show four spots separated azimuthally by 90° at a tilt angle of 35°, corresponding to the angle between (211) and (100) planes, and eight spots that are divided into two identical rectangular patterns, indicating the fourfold symmetry of the (100) plane. It is noted that the (211) planes are unique for simple-cubic perovskite MAPbBr$_3$ and do not present in the rock salt structure of PbS or the zinc-blende GaAs. The (211) patterns of MAPbBr$_3$ rotate by 45° in plane in relation to GaAs [100] direction, indicating the orientation of the MAPbBr$_3$ crystal. The clearly distinct spot patterns with a low background noise reveal the high quality of monocrystalline PbS and MAPbBr$_3$ films. Results of reciprocal space mapping (RSM) analysis are obtained around (300) and (331) points of MAPbBr$_3$ (Figure 2g). The asymmetric feature suggests the existence of different in-plane strains in different directions. Collectively, these experimental results clearly demonstrate the single crystal feature of the MAPbBr$_3$ film epitaxially formed on the PbS/GaAs substrate.

Monocrystalline GaAs (100) substrates have been extensively used in optoelectronic industry at affordable costs up to 8 inches in diameter[36]. Its widespread availability promises the potential to scale up the heteroepitaxy of MAPbBr$_3$ films. Figure 3a presents a MAPbBr$_3$/PbS/GaAs sample with a size of 1.5 × 5 cm$^2$, in which three different locations are independently examined with SEM and XRD. Similar surface morphologies obtained at these locations indicate that the sample has a very uniform MAPbBr$_3$ coating (Figure 3b), and consistent XRD peaks of the (211) plane during in-plane phi scanning reveal the single-crystal phase forming across the entire sample surface.

Based on the XRD results obtained in Figure 2, we establish an atomic model to interpret the MAPbBr$_3$/PbS/GaAs epi-structure, illustrated in Figure 4. The structure data is also provided as a supplemental file. According to Figures 2d–2f, the epitaxial relation is MAPbBr$_3$ (100) ∥ PbS (110) ∥ GaAs (100), MAPbBr$_3$ [001] ∥ PbS [001] ∥ GaAs [011], and MAPbBr$_3$ [010] ∥ PbS [1$\bar{1}$0] ∥ GaAs [01$\bar{1}$]. Specifically, the PbS layer grows up along its [110] direction on GaAs (100), and in-plane directions [001] and [1$\bar{1}$0] are parallel to GaAs [011] and [01$\bar{1}$], respectively. This observation is consistent with previous reports[37]. While there is only a ~5% mismatch between PbS [1$\bar{1}$0] and GaAs [01$\bar{1}$], the mismatch between PbS [001] and GaAs [011] is as large as ~50%. Therefore, the lattice registry is most likely metamorphic along GaAs [01$\bar{1}$] and coincident along GaAs [011], forming a superlattice with 3 GaAs and 2 PbS. Similar observations are obtained at the MAPbBr$_3$/PbS heterointerface. The lattice is almost perfectly matched between MAPbBr$_3$ [001] and PbS [001], but there is a ~50% mismatch between MAPbBr$_3$ [010] and PbS [1$\bar{1}$0]. Therefore, the coincident growth along PbS [1$\bar{1}$0] creates a superlattice with 3 PbS and 2 MAPbBr$_3$. Mechanisms for such a growth preference remain unclear, which are probably ascribed to the different crystal structures of GaAs (zinc blende), PbS (rock salt) and MAPbBr$_3$ (perovskite).

The epitaxial process for MAPbBr$_3$ on PbS/GaAs is further investigated by examining the film morphology with SEM and atomic force microscopy (AFM), at different time courses during the growth (Figure 5). The as-grown PbS/GaAs sample has a relatively smooth surface, with a root-mean-square (RMS) roughness of ~5.4 nm (Figure 5a). Observed surface defects are likely related to misfit dislocations. Clearly, the growth of MAPbBr$_3$ on PbS/GaAs divides into two distinct steps. In the first step, the solid-gas reaction of MABr and PbS occurs on the PbS surface. After CVD growth for 0.2 hours, a MAPbBr$_3$ film appears and forms an undulating surface with a period of ~70 nm and a roughness of ~8.5 nm (Figure 5b). In particular, the striped structures follow the cleavage directions of GaAs (100) wafers, which are [110] and [1$\bar{1}$0]. Such an anisotropic growth can be explained by the lattice matching conditions between MAPbBr$_3$ and PbS, which results in distinct appearance along the two in-plane directions [001] and [010] for MAPbBr$_3$ (Figure 4). The mechanisms are analogous to the classical Frank–van der Merwe growth in the MAPbBr$_3$ [001] direction, and the Volmer–Weber growth in the MAPbBr$_3$ [010] direction[38]. As the process continues, the MABr gas has to diffuse through the MAPbBr$_3$ film and reacts with the PbS underneath it. The second step forms more rectangular or cubic shaped MAPbBr$_3$ crystals and eventually results in a continuous film surface with terrace morphology (Figures 5c and 5d). After CVD growth for 8 hours, the MAPbBr$_3$ film has a roughness of ~12.6 nm. It is noted that the MAPbBr$_3$ growth rate considerably decreases as the film becomes thicker, since the film growth is eventually limited by the diffusion of MABr gas into the MAPbBr$_3$ film. Such growth kinetics is analogous to the thermal oxidation of silicon[39].

In summary, we demonstrate the heteroepitaxy of centimeter-scale, monocrystalline MAPbBr$_3$ films on GaAs (100) via the gas-solid reaction, with a solution-processed PbS interfacial layer. Such a growth strategy overcomes the limitations of previously explored methods, and heterogeneously integrates an emerging lead halide perovskite material with a traditional III–V compound semiconductor, in a single crystal format. The applied CBD and CVD processes, as well as the use of GaAs (100) wafers, are compatible with large-scale manufacturing. It is envisioned that these concepts could immediately realize high-quality monocrystalline perovskites on 6-inch and even 8-inch GaAs wafers. Future possibilities involve the large-area production of many other halide perovskites with similar lattice structures, for example, MAPbCl$_3$ ($a$ = 5.67 Å), CsPbCl$_3$ ($a$ = 5.61 Å) and CsPbBr$_3$ ($a$ = 5.87 Å). These monocrystalline materials could potentially produce perovskite-based optoelectronic devices with higher performance, although additional issues associated with device design and fabrications must be considered. Collectively, the concepts provide a promising direction for fundamental materials research and advanced device applications.

**Methods**

*PbS growth*

PbS films are grown on GaAs (100) wafers (single-side polished, updoped, AXT Inc.) via chemical bath deposition (CBD) at 50 °C in ambient environment (Figure 1a). The chemical solution contains 270 mM sodium hydroxide (NaOH powder, 97%, Alfa Aesar), 10 mM lead nitrate ($Pb(NO_3)_2$ powder, 99.99%, Aladdin) and 50 mM thiourea ($SC(NH_2)_2$ powder, 99%, Sigma Aldrich). The growth rate is ~1 nm/s. The polished side of GaAs faces down in the solution, to avoid PbS cluster particles randomly falling on the growth surface. The PbS deposition is also performed on Si (100) wafers and glass substrates for comparison.

*$MAPbBr_3$ growth on PbS/GaAs*

$MAPbBr_3$ films are formed on PbS/GaAs substrates via chemical vapor deposition (CVD) with a solid-gas reaction (Figure 1b). As-grown PbS/GaAs samples and excessive methylammonium hydrobromide (MABr powder, 98%, TCI Chemicals) are placed in a quartz tube furnace (OTF-1200X-S, tube diameter 50 mm, MTI Corp.) The reaction of MABr and PbS to form $MAPbBr_3$ occurs at 145 °C in argon environment at a pressure of ~100 Pa. The $MAPbBr_3$ growth is also performed on PbS/Si and PbS/glass substrates for comparison.

*$MAPbBr_3$ growth on Si and GaAs*

$MAPbBr_3$ growth is also attempted on Si and GaAs substrates via traditional CVD methods[28] without the PbS interlayer (Figure S1a). Excessive lead bromide ($PbBr_2$ powder, 99%, TCI Chemicals) and MABr powders are placed in the furnace at 350 °C in the upper stream of Ar flow (30 sccm, 100 Pa), while Si (100) or GaAs (100) wafers are placed in the down stream. The process lasts for ~20 mins.

*Materials Characterizations*

X-ray diffraction (XRD) patterns are obtained with a Rigaku S2 diffractometer (40 kV, 40 mA). Phi scanning are performed using constant $2\theta$ angles corresponding to the MAPbBr$_3$ (211) diffraction plane. XRD pole figures are measured using constant $2\theta$ angles corresponding to the GaAs/PbS (111) planes and the MAPbBr$_3$ (211) plane. Asymmetric high-resolution reciprocal space maps (HR RSM) around (300) and (133) points of MAPbBr$_3$ are plotted with the scanning mode on a Bruker D8 Discover. SEM images are taken with a Zeiss GeminiSEM 500 (15 kV), and EBSD data are measured with a Zeiss Merlin. AFM images are obtained with a BRUKER Multimode 8 (peak-force tapping mode). Atomic structures are established with Materials Studio 8.0 (Accelrys).

**Acknowledgements**

This work is supported by the Tsinghua University Initiative Scientific Research Program, the State Key Laboratory of New Ceramic and Fine Processing Tsinghua University (No. KF202108), Beijing Municipal Natural Science Foundation (4202032), National Natural Science Foundation of China (NSFC) (52171239, T2122010).

**Author contributions**

X. S. and D. K. developed the concepts. D. K., E. W., K. Z., and H. W. performed material growth and characterization. D. K., Y. Z., D. C, and X. S. performed modeling and simulations. K. L., L. Y. and X. S. supervised the research. D. K. and X. S. wrote the paper in consultation with other authors.

**Competing interests**

The authors declare no competing interests.

**Data and materials availability**

All data needed to evaluate the conclusions in the paper are present in the paper and/or the Supplementary information.

# Figure 1

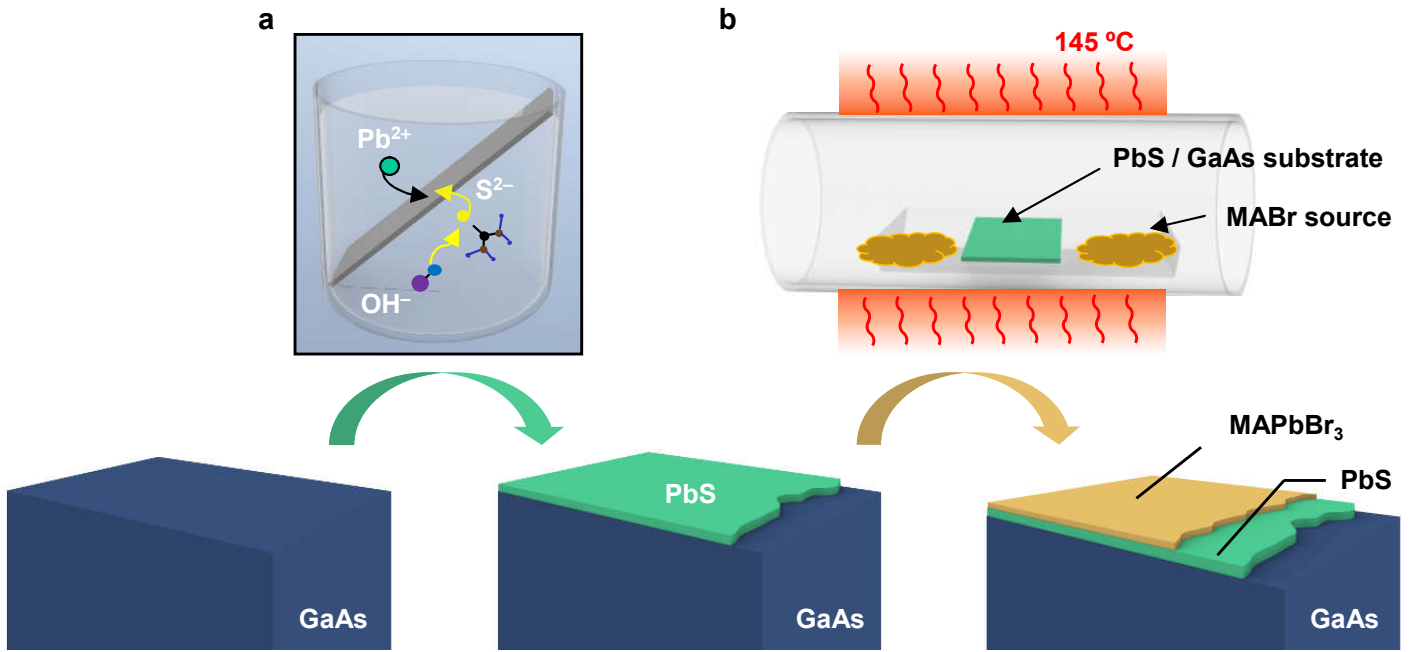

**Figure 1. Schematic illustration of the MAPbBr$_3$ epitaxy on GaAs.** (a) A single crystalline PbS film is grown on GaAs via chemical bath deposition. (b) The PbS film then reacts with MABr in a chemical vapor deposition tube to form a single crystalline MAPbBr$_3$ film.

# Figure 2

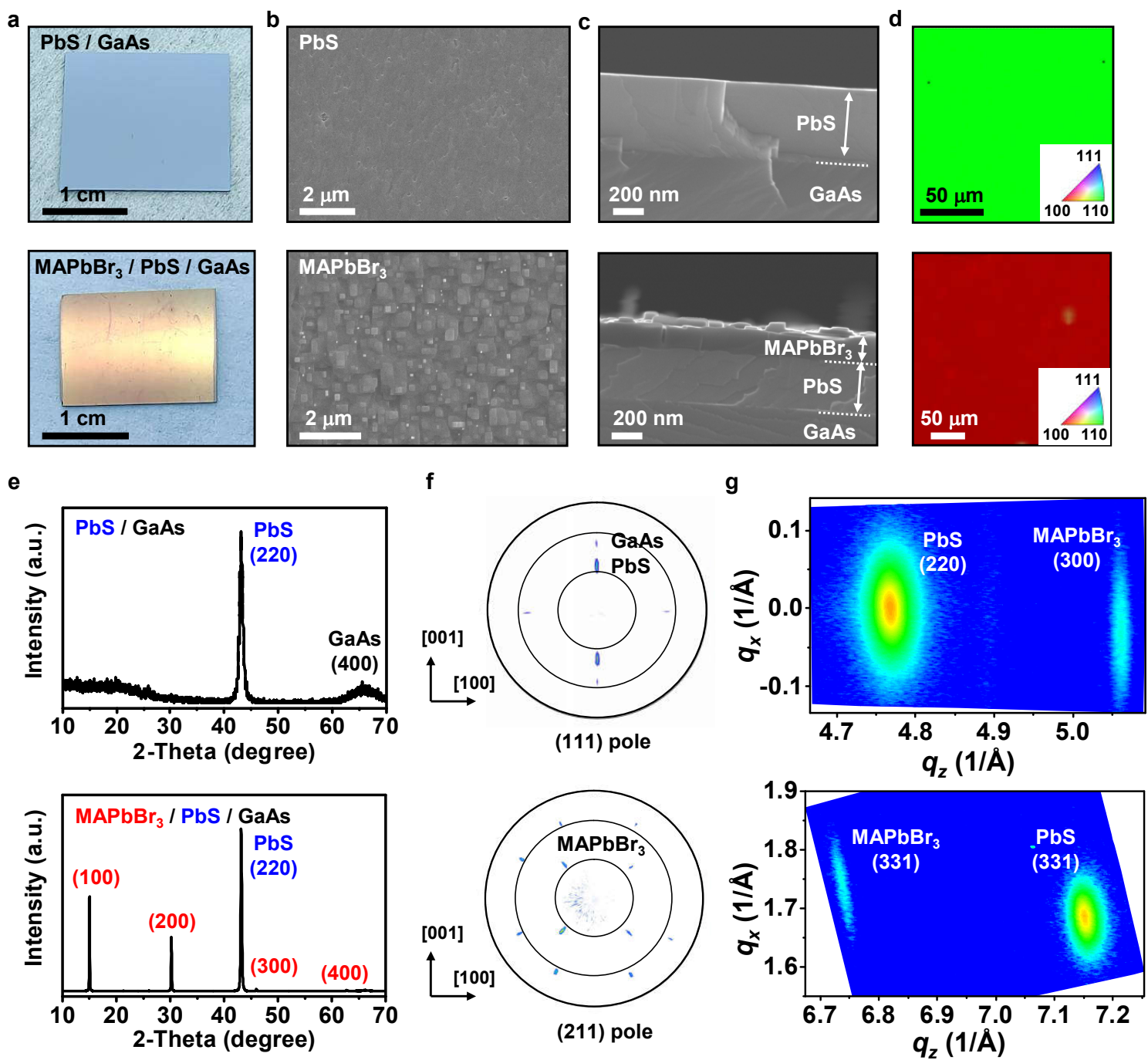

**Figure 2. Structural characterizations for PbS-on-GaAs and MAPbBr$_3$-on-PbS-on-GaAs samples.** (a) Top-view photographs, (b) Top-view SEM images, (c) Cross-sectional SEM images, (d) EBSD maps, (e) XRD patterns, and (f) Pole figures for two samples (top: PbS/GaAs; bottom: MAPbBr$_3$/PbS/GaAs). (g) RSMs around (300) and (331) points of MAPbBr$_3$.

**Figure 3**

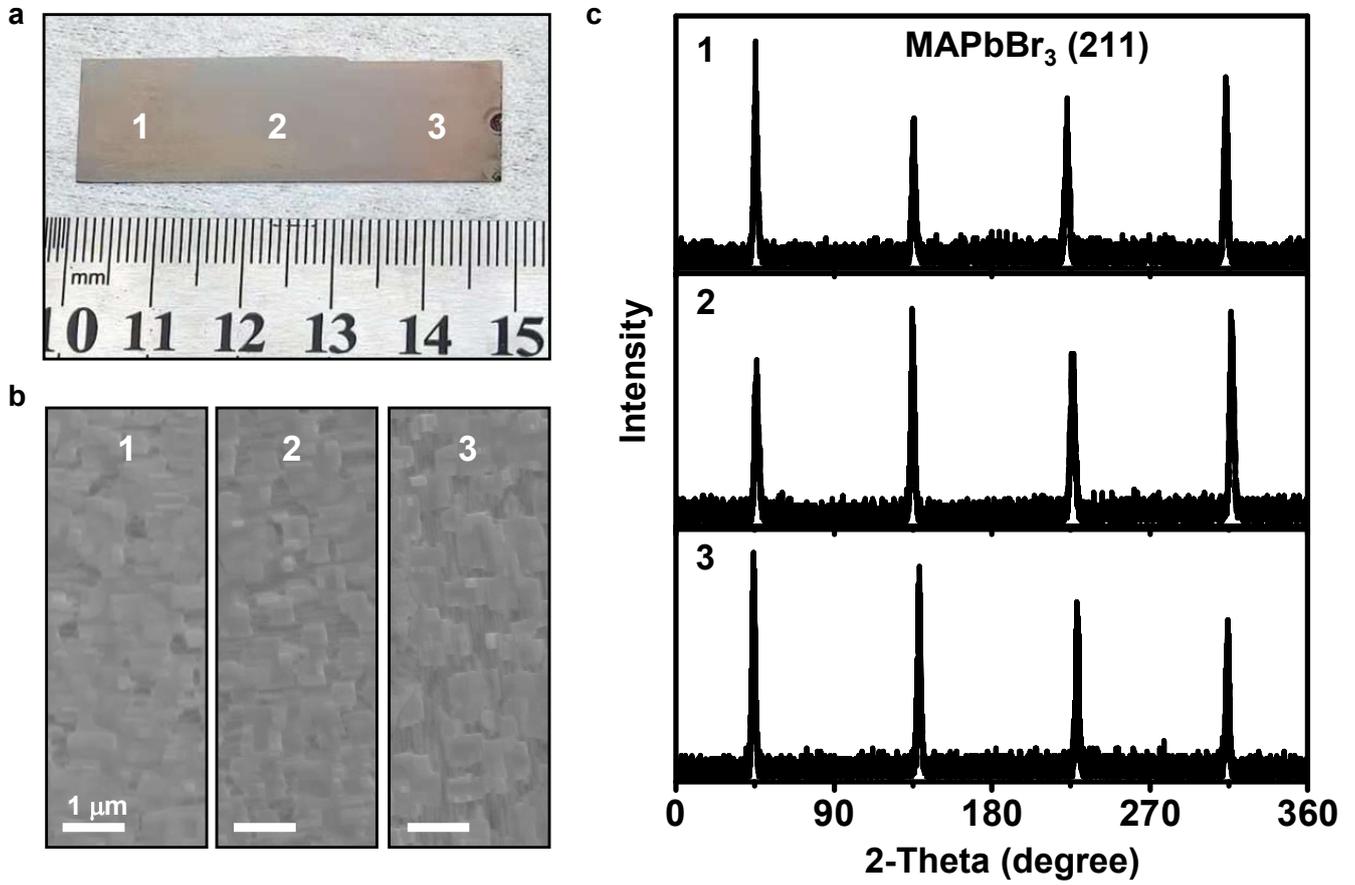

**Figure 3. A large-area single crystalline MAPbBr$_3$ film on PbS/GaAs.** (a) Photograph of the sample with a dimension of 1.5 × 5 cm$^2$. (b) Top-view SEM images at 3 different locations. (c) XRD phi-scan patterns of the MAPbBr$_3$ (211) plane at 3 different locations.

**Figure 4**

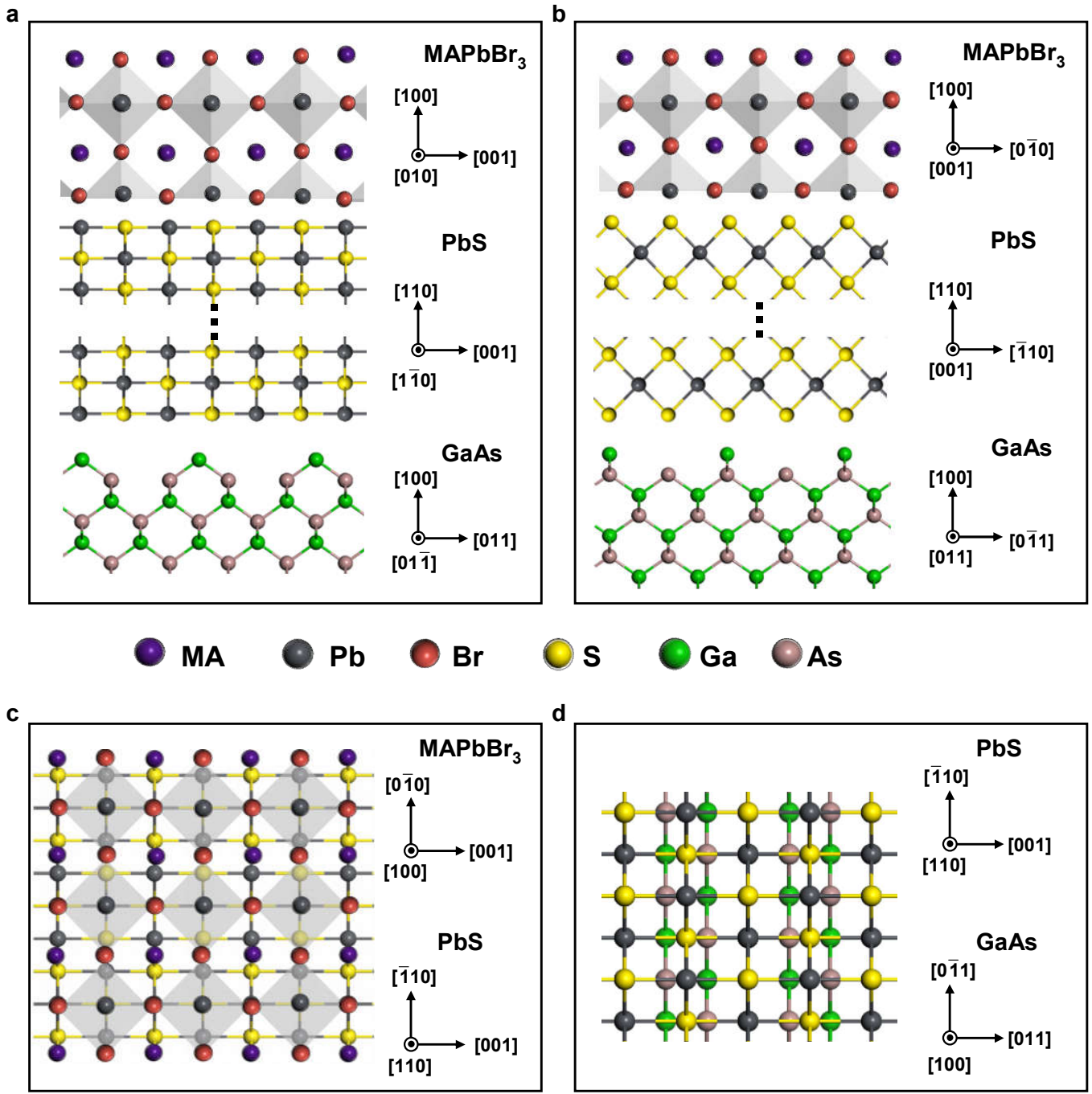

**Figure 4. Atomic model of the MAPbBr$_3$/PbS/GaAs epitaxial structure.** (a) Side view normal to the PbS [110] direction. (b) Side view normal to the PbS [001] direction. (c) Top view of the stacked MAPbBr$_3$ (100) / PbS (220) heterointerface. (d) Top view of the stacked PbS (220) / GaAs (100) heterointerface.

# Figure 5

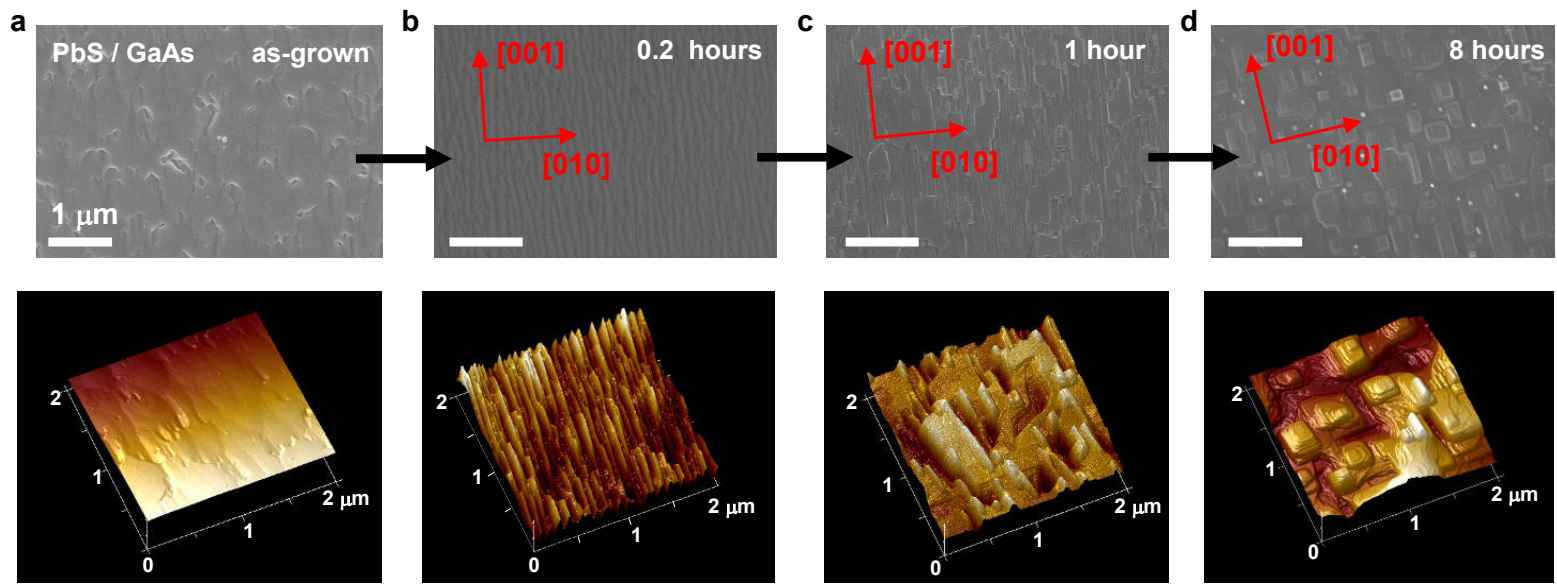

**Figure 5. Evolution of surface morphology during the MAPbBr$_3$ epitaxy on PbS/GaAs.** SEM (top) and AFM (bottom) images for (a) the as-grown PbS/GaAs sample, and after growth for (b) 0.2 hours, (c) 1 hour and (d) 8 hours. Arrows indicate the crystal directions of MAPbBr$_3$.

# Figure S1

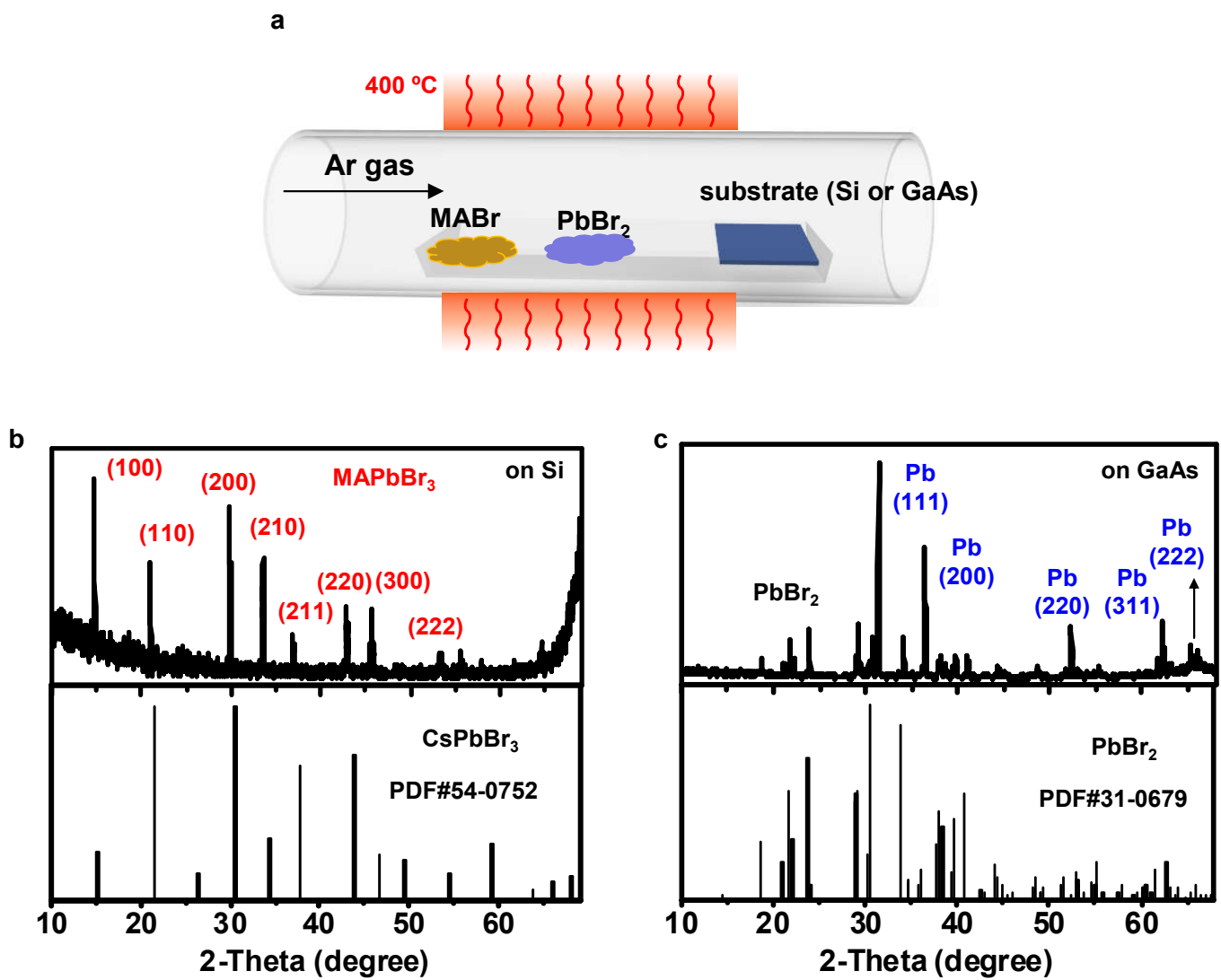

**Figure S1.** (a) MAPbBr$_3$ growth via conventional CVD, by reacting MABr and PbBr$_2$ on Si or GaAs at 400 ºC in argon environment. (b) XRD pattern for the CVD deposited film on Si, indicating the formation of polycrystalline MAPbBr$_3$. (c) XRD pattern for the CVD deposited film on GaAs, indicating the formation of polycrystalline PbBr$_2$ and Pb, but no MAPbBr$_3$. Powder Diffraction Files (PDF) of CsPbBr$_3$ and PbBr$_2$ are provided for comparison.

# Figure S2

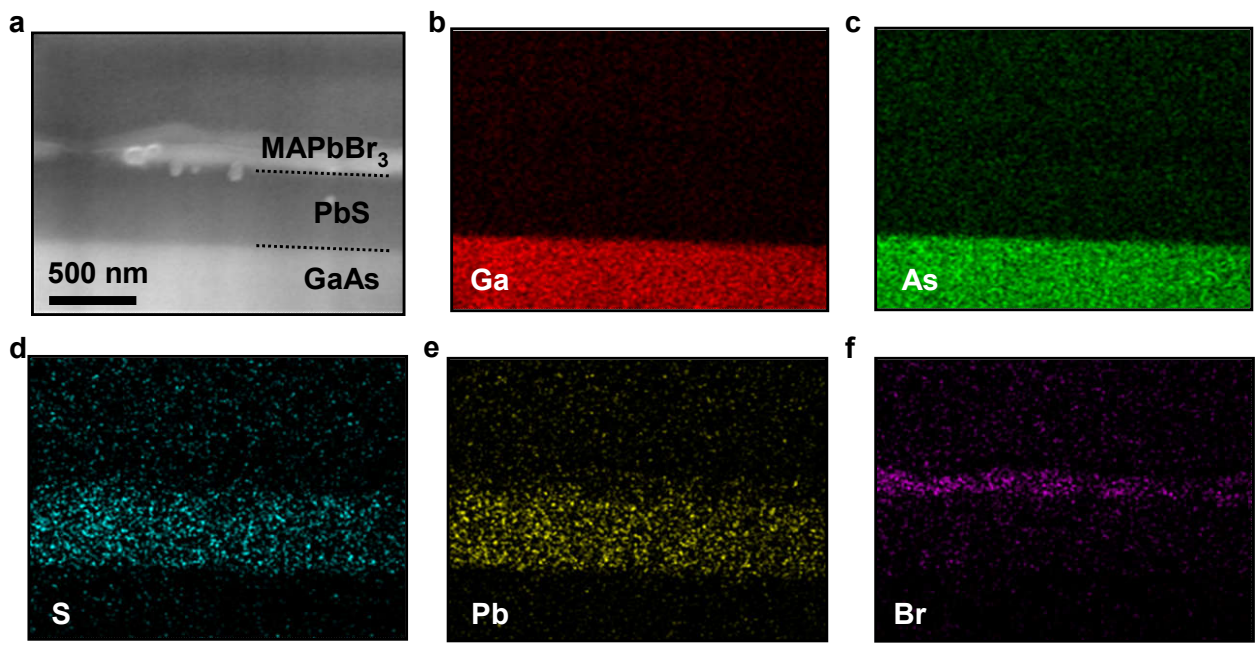

**Figure S2.** (a) Cross-sectional SEM image and (b–f) corresponding energy-dispersive X-ray spectroscopy (EDS) element mappings for the MAPbBr$_3$/PbS/GaAs sample.

# Figure S3

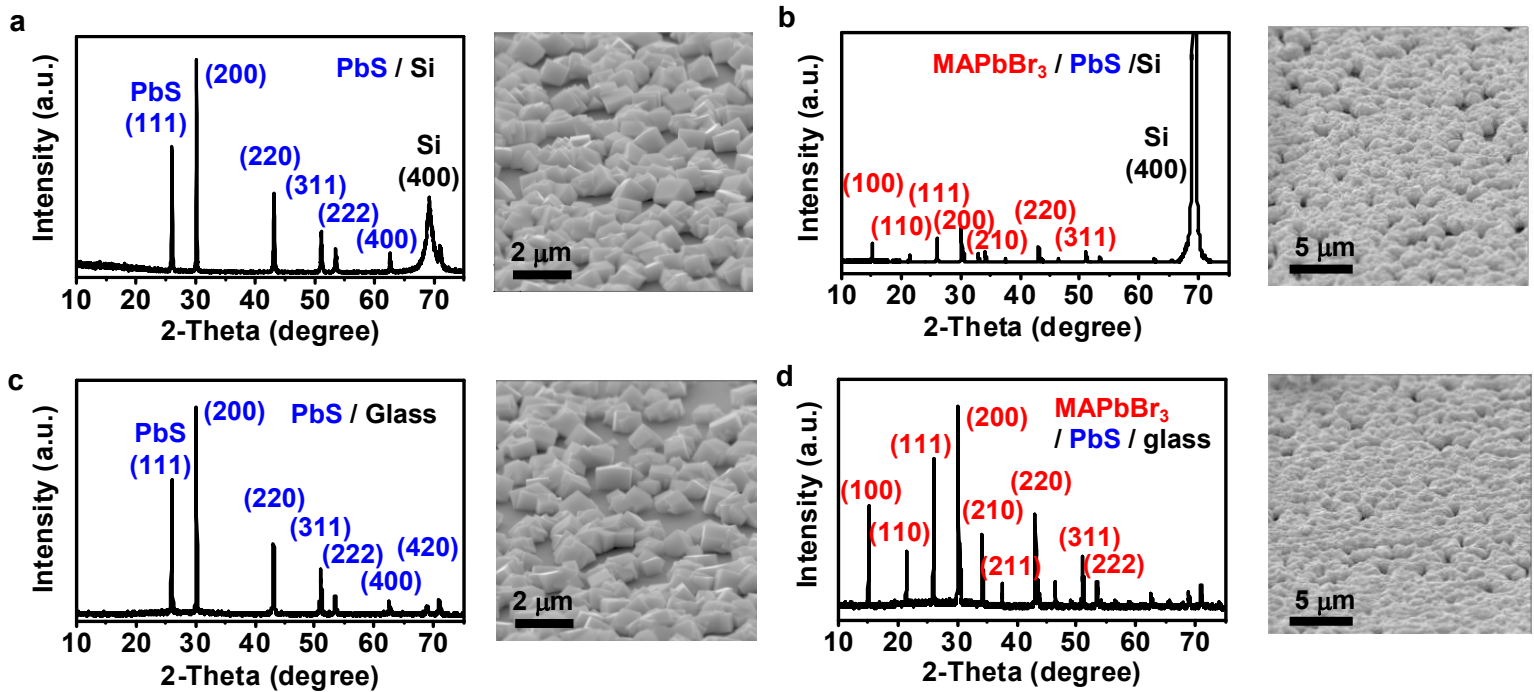

**Figure S3. Structural characterizations of PbS and MAPbBr₃ films grown on Si and glass substrates via the same process as in Figure 1.** XRD patterns (left) and SEM images (right) for (a) PbS on Si, (b) MAPbBr₃ on PbS/Si, (c) PbS on glass, (d) MAPbBr₃ on PbS/glass. All these films are polycrystalline.